\documentclass[12pt]{article}
\usepackage{amsmath,amssymb,graphicx,epsfig}
\def\bea{\begin{eqnarray}}
\def\eea{\end{eqnarray}}
\def\be{\begin{equation}}
\def\ee{\end{equation}}
\def\ba{\begin{array}}
\def\ea{\end{array}}

\def\Z{{\bf Z}}

\def\bari{i \hspace{-3pt} \bar{\;\raisebox{3.7pt}{}}}
\def\barj{j \hspace{-3pt} \bar{\;\raisebox{3.7pt}{}}}

\font\tenrsfs=rsfs10
\font\sevenrsfs=rsfs7
\font\fiversfs=rsfs5
\newfam\rsfsfam
\textfont\rsfsfam=\tenrsfs
\scriptfont\rsfsfam=\sevenrsfs
\scriptscriptfont\rsfsfam=\fiversfs
\def\mathscr#1{{\fam\rsfsfam\relax#1}}

\begin{document}
\thispagestyle{empty}
\begin{center}
\begin{small}
\hfill NEIP/06-05
\end{small}
\begin{center}

\vspace{1.5cm}

{\Large \bf Constraints for the existence of flat and stable \\[3mm] 
non-supersymmetric vacua in supergravity}

\end{center}

\vspace{1cm}

{\large \bf M.~G\'omez--Reino} {\large and} {\large \bf C.~A.~Scrucca} \\[2mm]

\vspace{7mm}

{\em Institut de Physique, Universit\'e de Neuch\^atel,\\ 
Rue Breguet 1, CH-2000 Neuch\^atel, Switzerland}
\vspace{.2cm}

\end{center}

\vspace{0.4cm}
\centerline{\bf Abstract}
\vspace{-0.1cm}
\begin{quote}

We further develop on the study of the conditions for the existence of locally stable non-supersymmetric
vacua with vanishing cosmological constant in supergravity models involving only chiral superfields.
Starting from the two necessary conditions for flatness and stability derived in a previous paper
(which involve the K\"ahler metric and its Riemann tensor contracted with the supersymmetry breaking
auxiliary fields) we show that the implications of these constraints can be worked out exactly not only 
for factorizable scalar manifolds, but also for symmetric coset manifolds. In both
cases, the conditions imply a strong restriction on the K\"ahler geometry and constrain the vector of
auxiliary fields defining the Goldstino direction to lie in a certain cone.
We then apply these results to the various homogeneous coset manifolds spanned by the
moduli and untwisted matter fields arising in string compactifications, and discuss their implications.
Finally, we also discuss what can be said for completely arbitrary scalar manifolds, and derive in this
more general case some explicit but weaker restrictions on the K\"ahler geometry.

\vspace{5pt}
\end{quote}

\newpage

\renewcommand{\theequation}{\thesection.\arabic{equation}}

\section{Introduction}
\setcounter{equation}{0}

Supergravity theories represent one of the phenomenologically most promising class of supersymmetric 
extensions of the standard model \cite{gravmed,gravmedunif}. They are also very well motivated at the 
theoretical level, since they can emerge as low-energy effective theories of string models. 
Furthermore, the moduli fields arising in string compactifications to four dimensions seem to be
natural candidates to constitute the hidden sector that is supposed to be responsible
for supersymmetry breaking \cite{dilatondom,bimsoft}. From a phenomenological point of view, this type 
of models must however posses some characteristics in order to be viable: supersymmetry must be 
broken, the cosmological constant should be tiny, and all the extra moduli fields of the hidden sector 
should be stabilized with a sufficiently large mass. In the low energy effective theory all these crucial 
features are controlled by a single quantity, the four-dimensional scalar potential, which gives 
information on the dynamics of the moduli fields, on how supersymmetry is broken and on
the value of the cosmological constant. The characterization of the conditions under which a
supersymmetry-breaking stationary point of the scalar potential satisfies simultaneously the flatness
condition (vanishing of the cosmological constant) and the stability condition (the stationary point is
indeed a minimum) is therefore very relevant in the search of phenomenologically viable string models.

Recently, some substantial progress has been achieved in the search of these non-supersymmetric
Minkowski/dS vacua in the context of string/M-theory compactifications. This was mainly related 
to the understanding of the superpotentials generated by background fluxes \cite{GVW} and by 
non-perturbative effects like gaugino condensation \cite{GC}, which suggested then new 
interesting possibilities for model building, like in particular those proposed in refs.~\cite{KKLT, GKP}. 
On the other hand, it is interesting to note that the structure of the K\"ahler potential is usually fixed 
by the symmetries of the compactification, whereas the form of the superpotential is more difficult to 
characterize. Therefore any information allowing for a discrimination among compactifications with 
different K\"ahler potentials, independently of the form of the superpotential, is therefore extremely 
valuable. One could then hope to eventually combine all these sources of information to try to identify 
a hopefully restricted set of phenomenologically viable models.

In this respect it was shown by the authors in ref.~\cite{us} that, in models involving an arbitrary 
number of chiral multiplets but no vector multiplets, it is possible to derive two very simple and 
strong conditions for the existence of flat and stable vacua, which are necessary but in general
not sufficient. These conditions involve the K\"ahler metric and its Riemann tensor as well as 
the vector of auxiliary fields controlling the direction of supersymmetry breaking. One can then 
imagine a situation with a fixed K\"ahler potential and an arbitrary superpotential and study the 
space of solutions admitted by these constraints by scanning over all the possible values of the 
vector of auxiliary fields satisfying the restrictions with a fixed K\"ahler metric and 
Riemann tensor. This constrains the K\"ahler geometry of the scalar manifold. 
Explicit expressions for these constraints were derived in \cite{us} in the case where the K\"ahler
manifold spanned by the scalar fields is a factorizable space. In that case, the resulting conditions 
restrict the K\"ahler curvature scalars associated to each of the scalar fields and the ratios of the 
supersymmetry-breaking auxiliary fields defining the Goldstino direction. These results were then 
applied to the dynamics of moduli fields arising in some string compactifications.

The aim of this paper is to further develop on the study initiated in \cite{us} and analyze in more
generality the implications that the flatness and stability constraints have on theories with more
general scalar manifolds. We will study in detail the class of theories where the scalar manifold
is not factorizable but has instead the special feature of being symmetric, as is the case for instance 
for the coset spaces that are relevant for string models. Actually in this case of symmetric 
manifolds it is possible to derive, as in the case of factorizable manifolds, very simple and 
strong constraints. We will also study what can be said in the general case where the scalar 
manifold is completely arbitrary.

The paper is organized as follows: In Section 2 we review the derivation of the necessary conditions 
for flatness and stability of ref.~\cite{us}, which represent the starting point of our analysis. In Section 3 
we examine what kind of implications can be extracted from these constraints for completely general 
scalar manifolds. In Sections 4 and 5 we study instead the special cases of factorizable and symmetric 
scalar manifolds, where more specific information restricting the K\"ahler geometry can be obtained 
and we also derive the bounds on the values that the auxiliary fields can take. In Section 6 we apply 
our results to the particularly interesting case of homogeneous coset spaces arising in the moduli 
sector of string compactifications. In Section 7 we also comment on the implications of our results for 
supersymmetric Randall--Sundrum  models. Finally in Section 8 we summarize our results.

\section{Conditions for flat and stable vacua}
\setcounter{equation}{0}

The Lagrangian of a minimal four-dimensional supergravity theory with $N$ chiral superfields is 
entirely specified, at the leading two-derivative order, by a single arbitrary real function $G$ depending 
on the corresponding chiral superfields $\Phi_i$ and their conjugates $\Phi_i^\dagger$, as well as on 
its derivatives \cite{sugra}. The function $G$ is the following K\"ahler invariant combination of the real 
K\"ahler potential $K$ and the holomorphic superpotential $W$ (we use Planck units $M_{\rm P}=1$):
\be
G(\Phi_i,\Phi_i^\dagger) = K(\Phi_i,\Phi_i^\dagger) + \log W(\Phi_i) +
\log \bar W(\Phi_i^\dagger)\,.
\ee
Mixed holomorphic and antiholomorphic derivatives of $G$ depend only on the K\"ahler potential $K$ 
and define the K\"ahler geometry of the manifold parametrized by the scalar fields, whose metric is given 
by $g_{i \barj} = G_{i \barj}$. Purely holomorphic or antiholomorphic derivatives of $G$ depend instead 
also on the superpotential $W$, and control the way supersymmetry is broken. The auxiliary fields $F^i$ 
of the chiral multiplets are determined by their equations of motion to have the values $F^i = e^{G/2} G^i$.
The potential for the scalar fields $\phi^i$ takes then the following simple form:
\be
\label{genpot}
V = e^G \left(G^i G_{i}-3\right) \,.
\ee

In order to study the existence of non-supersymmetric Minkowski minima in a potential of the type (\ref{genpot}) 
one should first check under which conditions this potential has a stationary point with vanishing cosmological 
constant. The flatness condition of vanishing cosmological constant implies that $V=0$ at the minimum, implying:
\be
g_{i \barj}\, G^i G^{\barj} = 3 \,.
\label{flatnessbis}
\ee
The stationarity condition implies instead that $\nabla_i V = 0$ at the vacuum, which leads to the following 
equations:
\be
\label{4}
G_i + G^k \nabla_i G_k = 0 \,,
\ee
where $\nabla_i$ denotes the covariant derivative on the K\"ahler manifold, which when applied to $G_i$ 
gives $\nabla_i G_k = g_{ik}-G_{ij\bar l}\,G^{\bar l}$. 

Finally, to ensure the stability of the stationary point, one should check that the matrix of second derivatives 
of the potential is positive definite \footnote{Notice that we require that the cosmological constant should be 
tunable to zero order by order in perturbation theory, and that there should not be any flat direction. 
Models of the no-scale type \cite{noscale} are thus exclude from our analysis.}. 
This matrix can be also computed using covariant derivatives, since the 
extra connection terms vanish by the flatness and stationarity conditions.
There are two different $n$-dimensional blocks, $V_{i \barj} = \nabla_i \nabla_{\barj} V$ and
$V_{i j} = \nabla_i \nabla_j V$, and after a straightforward computation (see also ref.~\cite{supertrace,FKZ}) 
these are found to be given by the following expressions:
\bea
\label{vij}
\begin{array}{lll}
V_{i \barj} \!\!\!&=&\!\!\! e^G\,\Big(g_{i \barj} + \nabla_i G_k \nabla_{\barj} G^k 
- R_{i \barj p \bar q}\, G^p\, G^{\bar q} \Big)\,, \smallskip\ \\
V_{i j} \!\!\!&=&\!\!\! e^G\,\Big(\displaystyle{\nabla_i G_j + \nabla_j G_i 
+ \frac 12 G^k \big\{\nabla_i,\nabla_j \big\}G_k}\Big) \,, \\
\end{array}
\eea
where $R_{i\barj p\bar q}$ denotes the Riemann tensor on the K\"ahler manifold, whose components are 
given by $R_{i\barj p\bar q}=G_{i\barj p\bar q}-g^{k\bar l}\,G_{i p\bar l}\,G_{\barj \bar q k}$.

The full $2n$-dimensional matrix of second derivatives at the stationary point has the form:
\begin{equation}
V_{IJ} = \left(
\begin{matrix}
V_{i \barj} &  V_{ij} \smallskip \cr
V_{\bari \barj} & V_{\bari j}
\end{matrix}
\right) \,.
\label{VIJ}
\end{equation}
The conditions under which this $2N$-dimensional matrix is positive definite are difficult to work out 
in general \footnote{See ref.~\cite{lukas} for an attempt in this direction for string models with fluxes.}, 
the only way being to study in full detail the behavior of all the $2N$ eigenvalues. However, 
it was shown in \cite{us} that it is possible to deduce some simple necessary conditions for the matrix 
(\ref{VIJ}) to be positive definite. This is done by using the property that if a matrix is positive definite 
then all its upper-left submatrices are also positive definite. This implies, for instance, that the 
$N$-dimensional submatrix $V_{i \barj}$ should be positive definite. In particular along the 
direction in the scalar field space defined by $G^i$ \footnote{Notice that in the fermion field space the 
direction defined by $G^i$ identifies the would-be Goldstino that is absorbed by the gravitino field 
through the super-Higgs mechanism when supersymmetry is broken.} one must therefore have 
$V_{i \barj} G^i {\bar G}^{\barj}>0$. It is straightforward to show that, using the flatness and stationarity 
conditions as well as the results (\ref{vij}), this yields to the extremely simple necessary conditions 
for stability \footnote{This condition can be easily generalized to the case 
in which a positive cosmological constant is required, with $V = e^G \, \epsilon$ at the vacuum.
One finds in that case $R_{i \barj p \bar q}\, G^i G^{\barj} G^p G^{\bar q} < 6+2\epsilon$.}:
\be
R_{i \barj p \bar q}\, G^i G^{\barj} G^p G^{\bar q} < 6 \;.
\label{main}
\ee

Equations (\ref{flatnessbis}) and (\ref{main}) represent simple and very strong constraints that should be 
fulfilled by any theory for the existence of non-supersymmetric vacua that are flat and stable. 
It is important to realize that the metric $g_{i \barj}$ and the curvature tensor $R_{i \barj p \bar q}$ depend 
only on the K\"aler potential and therefore on the geometry. On the other hand, the quantities $G^i$ 
depend also on the superpotential and define the way in which supersymmetry is broken, since they 
are related to the auxiliary fields by the relation $G^i = F^i/m_{3/2}$, where $m_{3/2} = e^{G/2}$. 
One can then imagine a situation with a fixed K\"ahler potential and an arbitrary superpotential. 
More precisely, one can treat $g_{i \barj}$ and $R_{i \barj p \bar q}$ as fixed quantities and scan over 
all the possible values of $G^i$ satisfying the restriction (\ref{flatnessbis}) and the bound (\ref{main}). It is 
then clear that eq.~(\ref{main}) puts constraints on the values that the various $G^i$ can take, and actually 
requiring eq.~(\ref{main}) to have a solution also requires that $g_{i \barj}$ and $R_{i \barj p \bar q}$ satisfy 
certain conditions. 

Notice that the two conditions (\ref{flatnessbis}) and (\ref{main}) are evaluated at a specific stationary point,
determined by the equations (\ref{4}). It is then very convenient to switch to normal coordinates around 
the point under consideration. This can be done by introducing the vielbein $e^J_i$ and its inverse $e_I^j$, 
defined in the usual way to diagonalize the metric and its inverse: 
$g_{i \barj} = e^R_i e^{\bar S}_{\barj} \delta_{R \bar S}$, and 
$g^{i \barj} = e_R^i e_{\bar S}^{\barj} \, \delta^{R \bar S}$ (in what follows we will use capital letters 
to denote flat indices). The flatness and stability conditions (\ref{flatnessbis}) and (\ref{main}) can then be
rewritten in terms of the metric $\delta_{I \bar J} = e_I^r e_{\bar J}^{\bar s} \, g_{r \bar s}$ and the Riemann 
tensor $R_{I \bar J P \bar Q} = e_I^r e_{\bar J}^{\bar s}  e_P^t e_{\bar Q}^{\bar u} \, R_{r \bar s t \bar u}$ in 
these coordinate, and in terms of the corresponding new variables $G^I = e^I_r G^r$:
\begin{eqnarray}
\begin{array}{l}
\delta_{I \bar J} \, G^I G^{\bar J} = 3 \,, \\[2mm]
R_{I \bar J P \bar Q} \, G^I G^{\bar J} G^P G^{\bar Q} < 6 \,.
\end{array}
\label{condgen}
\end{eqnarray}
These expressions will be our starting point. They clearly represent a constraint on the curvature evaluated 
at the given stationary point and on the directions of supersymmetry breaking. The strength and the simplicity 
of these constraint depend on the type of scalar manifold. In particular, it is clear that for homogeneous 
manifolds with constant curvature they will translate into very direct constraints on the parameters of the theory.

Unfortunately, as the conditions (\ref{condgen}) are quadratic and quartic in the variables $G^i$, 
is not possible in general to solve such conditions exactly. To derive explicit results one must 
either further simplify the conditions to get a new set of weaker but still necessary conditions 
that can be solved exactly, or consider special types of geometries for which the problem simplifies from a 
complicated quartic problem to an exactly solvable quadratic problem. We will explore what can be said 
using both different options in the following sections.

\section{General scalar manifolds}
\setcounter{equation}{0}

In this section we want to explore the possibility to use the conditions (\ref{condgen}) to find a 
restriction on the K\"ahler geometry in a totally generic case where the manifold spanned by the scalar fields 
is an arbitrary K\"ahler manifold. In order to do that the key is to try to reduce the problem from a quartic one
to a quadratic one by defining new variables that are quadratic in the $N$ complex quantities $G^i$. In doing 
so we will unavoidably introduce new constraints, which keep the difficulty of the problem intact. However, 
we can then take the option of discarding the new constraints and solving the weaker set of conditions exactly
to obtain a general but weaker necessary condition. There are actually two different possible ways 
to do this, which lead to different conditions.

A first possibility that one can consider to try to solve the conditions (\ref{condgen}) is to introduce 
the $N$-dimensional positive definite Hermitian matrix of variables
\be
\label{herm}
H^{I \bar J} = \frac 13\, G^I G^{\bar J}\,.
\ee
Clearly, this does not represent a regular change of variables in terms of the $G^i$'s. 
Indeed, the $N^2$ real components of $H^{I \bar J}$ are subject to the following quadratic 
constraints:
\be
\label{lm1}
H^{I \bar J} H^{P \bar Q} = H^{I \bar Q} H^{P \bar J} \,.
\ee
These represent $(N-1)^2$ independent real constraints, leaving $2N-1$ real independent variables, 
which correspond to the $N$ absolute values and the $N-1$ relative phases of the variables $G^I$. 
The flatness and stability conditions (\ref{condgen}) can then be rewritten in 
terms of the new variables (\ref{herm}) as 
\be
\delta_{I \bar J}\,H^{I \bar J} = 1\;,\;\;
R_{I \bar J P \bar Q} \, H^{I \bar J} H^{P \bar Q} < \frac 23 \,.
\label{conda}
\ee
This is a constrained minimization problem which can be solved in the standard way using Lagrange 
multipliers. The difficulty now is that, in addition to the linear flatness condition, the variables 
$H^{I\bar J}$ are also subject to the quadratic constraints (\ref{lm1}). The treatment of these 
constraints with Lagrange multipliers implies cubic terms in the functional to be minimized, and 
therefore the problem cannot be solved exactly. Then, as we already mentioned, the best we can do is 
to discard the constraints $(\ref{lm1})$ and consider the weaker set of conditions defined by (\ref{conda})
on the variables $H^{I\bar J}$. 

The problem defined by the eqs.~(\ref{conda}) can be solved by considering the linear map 
${H^{I \bar J} \rightarrow R^{\bar J I}}_{\!\! P \bar Q} \,H^{P \bar Q}$ on Hermitian tensors. 
This map acts on the $N^2$-dimensional vector space of independent components of the Hermitian 
tensors $H^{I \bar J}$, and can be represented by a $N^2 \times N^2$ matrix. This matrix can then 
be diagonalized, and this defines $N^2$ eigenvalues $R_h$, with $h=1,2,\dots,N^2$. The corresponding 
eigenvectors $H_h^{I \bar J}$ satisfy the eigenvalue equations
\be\label{eigen1}
R^{\bar J  I}_{\;\;\; P \bar Q} \,H_h^{P \bar Q} = R_h\, H_h^{I \bar J} \,.
\ee

It is then clear from (\ref{eigen1}) that if any of the eigenvalues $R_h$ are negative or vanishing, then 
the constraints (\ref{conda}) admit solutions as long as the variables $H^{I \bar J}$ are aligned closely
enough to the particular directions associated to the negative or vanishing eigenvalues. On the other hand, 
if all the eigenvalues $R_h$ are positive, then the constraints (\ref{conda}) can be minimized 
straightforwardly using Lagrange multipliers. One finds that they admit solutions only if the following 
bound is satisfied:
\be
\delta^{I \bar J} \delta^{P \bar Q} \, R^{-1}_ {I \bar J P \bar Q} > \frac 32 \,.
\label{result1}
\ee

A second possibility that one can consider to try to solve the conditions (\ref{condgen}) 
is to introduce the $N$-dimensional complex symmetric matrix of variables
\be
\label{sym}
S^{I J} = \frac 13 \, G^I G^J\,.
\ee
Again, this does not represent a regular change of variables with respect to the variables $G^I$. 
Indeed, the $N(N+1)/2$ complex components of $S^{IJ}$ are subject to the following quadratic constraints:
\be
\label{lm2}
S^{I J} S^{P Q} = S^{I Q} \, S^{P J} \,.
\ee
These represent $N(N-1)/2$ independent complex constraints, so that one is left with $N$ complex 
independent variables, which are in one to one correspondence with the $N$ complex variables $G^I$. 

It is clear that, in order to be able to use efficiently this new set of variables, we need to squared the 
flatness condition. By doing so, the two flatness and stability conditions (\ref{condgen}) can be 
rewritten as:
\be
\delta_{I \bar J}\,\delta_{P \bar Q}\, S^{I P} S^{\bar J \bar Q} = 1\;,\;\; 
R_{I \bar J P \bar Q} \, S^{I P}  S^{\bar J \bar Q} < \frac 23 \,.
\label{f1}
\ee
This is again a constrained minimization problem that can be faced using Lagrange multipliers. 
But again there is the difficulty that, in addition to the flatness constraint, the variables $S^{IJ}$ 
are also subject to the quadratic constraints (\ref{lm2}). As before, the implementation of these
constraints with Lagrange multipliers implies cubic terms in the functional to be minimized, and
therefore the problem cannot be solved exactly either. 
Once again the best we can do is to discard the constraints 
(\ref{lm2}) and consider the weaker constraints represented only by the conditions (\ref{f1}).

The problem defined by the constraints (\ref{f1}) can be solved by considering the linear map 
$S^{I J} \rightarrow {R_{P \;\; Q}}^{\hspace{-10pt} I \;\;\, J} \,S^{P Q}$ on complex symmetric tensors. 
This map acts on the $N(N+1)/2$-dimensional vector space of independent 
components of the complex symmetric tensors $S^{I J}$, and can be represented by a 
$N(N+1)/2 \times N(N+1)/2$ matrix. This matrix can then be diagonalized, and this defines 
$N(N+1)/2$ eigenvalues $R_s$, with $s=1,2,\dots,N(N+1)/2$. 
The corresponding eigenvectors ${S_s}^{\! I J}$ satisfy the eigenvalue equations
\be
{R_{P \;\; Q}}^{\hspace{-10pt} I \;\;\, J} \,S_s^{P Q} = R_s\, S_s^{I J} \,.
\ee
They can be chosen to form an orthonormal and complete basis of the vector space, with 
${S_s}^{\! I J} S_{s'\, J I} = \delta_{ss'}$ and $\sum_s {S_s}^{\! I J} S_{s\,P Q} = \delta^I_P\, \delta^{J}_{Q}$.
The Riemann matrix and its inverse, whenever it exists, can then be written in the form
${R_{P \;\; Q}}^{\hspace{-10pt} I \;\;\, J} = \sum_s R_s \, {S_s}^{\! I J} S_{s\, P Q}$ and 
$R^{-1\;\,I \;\; J}_{\;\;P \;\; Q} = \sum_s {R_s}^{\!\! -1} {S_s}^{\! I J} S_{s\, P Q}$, and the 
new variables $S^{I J}$ can be decomposed as $S^{I J} = \sum_s S_s \,V_s^{I J}$. 
Using this, the conditions (\ref{f1}) can finally be rewritten as
\bea
\mbox{\large $\sum$}_s \, S_s^2 = 1 \;,\;\;
\mbox{\large $\sum$}_s \, R_s \, S_s^2  < \frac 23 \,.
\label{f3}
\eea
It is clear that these constraints admit solutions if any of the eigenvalues $R_s$ is negative 
or vanishes. If instead all the eigenvalues $R_s$ are positive, then from (\ref{f3}) we get that 
they have to fulfill the bound:
\be
{\rm min} \big\{R_s \big\} < \frac 23 \,.
\ee
This condition can also be rewritten as:
\be
{\rm max} \Big\{\mbox{eigenvalues}\,\big(R^{-1\;\,I \;\; J}_{\;\;P \;\; Q} \big)\Big\} > \frac 32 \,.
\label{result3}
\ee

The inequalities (\ref{result1}) and (\ref{result3}) represent two different constraints on the K\"ahler 
curvature that have to be necessarily satisfied in order for the theory to have the chance 
of admitting flat and stable non-supersymmetric vacua. They are valid in full generality for any 
supergravity theory, with an arbitrary scalar manifold, under the sole assumption that the effects 
due to vector multiplets can be neglected. However, as already mentioned, they contain less 
information than the original constraints (\ref{condgen}). To derive stronger conditions from 
(\ref{condgen}), without any loss of information, it is necessary to consider more specific classes of 
scalar manifolds where a more explicit knowledge of the constraints on the variables $H^{I\bar J}$ 
and/or $S^{IJ}$ is available.
This depends obviously on the details of the model, and therefore to get such an information 
one should perform a case by case analysis. We will however see in the following two sections 
that stronger constraints emerging directly from (\ref{condgen}) can be 
derived for the two classes of scalar manifolds that are respectively factorizable and symmetric.

\section{Factorizable scalar manifolds}
\setcounter{equation}{0}

A first situation in which the conditions (\ref{condgen}) can be solved exactly is when the scalar 
manifold is factorizable into a product of one-dimensional submanifolds associated to each of the 
fields (this case was already worked out in detail in \cite{us} but for completeness we will 
briefly review it here). For factorizable spaces the K\"ahler potential is separable into a sum of terms, 
each of them depending on a single chiral field. 
The K\"ahler metric becomes then diagonal and has only $N$ non-zero elements $g_{i \bari}$. 
The Riemann tensor is also completely diagonal, and has only $N$ non-vanishing 
components $R_{i \bari i \bari}$, which are related to the diagonal components of the metric 
through the curvature scalars $R_i$ of the one-dimensional submanifolds associated to each 
of the fields. In flat indices, one finds then that the Riemann tensor is given by:
\be
\label{Riemannfac}
R_{I \bar J P \bar Q} = 
\left\{\!
\begin{array}{l}
R_i \;,\;\; \mbox{if} \; I = J = P = Q \,, \\[3mm]
0 \;\;\;,\;\; \mbox{otherwise.} \\
\end{array}
\right. \,.
 \ee
This form of the Riemann tensor implies that both maps on Hermitian and symmetric tensors 
introduced in the previous section have vanishing eigenvalues. More precisely, the map on 
Hermitian tensors has $N$ non-vanishing eigenvalues given by the curvature scalars $R_i$, 
and $N(N-1)$ vanishing eigenvalues, so that the condition (\ref{conda}) is trivially satisfied. 
Similarly, the map on symmetric tensors has $N$ non-vanishing eigenvalues given by the 
curvature scalars $R_i$, and $N(N-1)/2$ vanishing eigenvalues, so the condition (\ref{f1}) 
is also trivially satisfied.

Nevertheless, as was shown in \cite{us}, in this case it is possible to derive explicit results 
directly from the constraints (\ref{condgen}), thanks to the particularly simple form (\ref{Riemannfac}) 
that the Riemann tensor takes, and get more restrictive necessary conditions than the one implied by 
(\ref{result1}) and (\ref{result3}). To do so, one needs to introduce the following $N$ real and 
positive variables parametrizing the Goldstino direction:
\be
\Theta_i = \frac 1{\sqrt{3}}\,\big|G^I\big| \,.
\ee
The two constraints (\ref{condgen}) become then
\be
\mbox{\large $\sum$}_i \Theta_i^2 = 1 \;,\;\; \mbox{\large $\sum$}_i R_i\, \Theta_i^4 < \frac 23 \;.
\ee
These can have solutions only if some of the curvature scalars are negative or vanish, or if they are 
all positive but satisfy the following bound:
\be
\mbox{\large $\sum$}_i R_i^{-1} > \frac 32 \,.
\label{constraintRn}
\ee
If the restriction (\ref{constraintRn}) is satisfied  then solutions exist, but only for a limited range 
of values for the variables $\Theta_i$. More precisely, the allowed interval is 
$\Theta_i \in [\Theta_{i}^-,\Theta_{i}^+]$, where:
\begin{eqnarray}
\begin{array}{lll}
\Theta_{i}^+ \!\!\!&=&\!\!\!
\left\{\begin{array}{lll}
\displaystyle{
\sqrt{\frac {R_i^{-1} \!+\! \sqrt{\displaystyle{\frac 23} R_i^{-1} \Big(\mbox{\large $\sum_{k \neq i}$}  R_k^{-1}\Big)
\Big(\mbox{\large $\sum_k$} R_k^{-1} \!-\! \displaystyle{\frac 32} \Big)}}{\Big(\mbox{\large $\sum_k$} R_k^{-1}\Big)}}} \,,
&\mbox{if}& R_i^{-1} < \displaystyle{\frac 32}\,, \hspace{-30pt} \bigskip\ \\[-2mm]
1\,, &\mbox{if}& R_i^{-1} > \displaystyle{\frac 32}\,. \hspace{-30pt} \smallskip\ \\
\end{array}\right. \bigskip\ \\
\Theta_{i}^- \!\!\!&=&\!\!\!
\left\{\begin{array}{lll}
\displaystyle{
\sqrt{\frac {R_i^{-1} \!-\! \sqrt{\displaystyle{\frac 23} R_i^{-1} \Big(\mbox{\large $\sum_{k \neq i}$}  R_k^{-1}\Big)
\Big(\mbox{\large $\sum_k$} R_k^{-1} \!-\! \displaystyle{\frac 32} \Big)}}{\Big(\mbox{\large $\sum_k$} R_k^{-1}\Big)}}} \,,
&\mbox{if}& \mbox{\large $\sum_{k \neq i}$} R_k^{-1} < \displaystyle{\frac 32}\,, \hspace{-30pt} \bigskip\ \\[-2mm]
0\,, &\mbox{if}&  \mbox{\large $\sum_{k \neq i}$} R_k^{-1} > \displaystyle{\frac 32}\,. \hspace{-30pt} \smallskip\ \\
\end{array}\right. \smallskip\ \\
\end{array}
\label{thetacritn}
\end{eqnarray}
This also constrains the values that the auxiliary fields can take, since these are given by 
$|F^I|=\sqrt{3}\, \Theta_i \, m_{3/2}$.

\section{Symmetric scalar manifolds}
\setcounter{equation}{0}

Another interesting and relevant case where one can solve the original constraints (\ref{condgen})
exactly, is when the K\"ahler manifold spanned by the scalar fields is a coset group manifold of the 
form $G/H$, where $G$ is the global isometry group and $H$ the local stability group. In this case 
the K\"ahler potential $K$ has a very special form due to the fact that the K\"ahler manifold has a 
large number of Killing vectors. These coset K\"ahler manifolds have been classified, and there 
exist finitely many types of them for each given dimensionality $N$ (see for example \cite{calabi}). 
All of them are Einstein manifolds and, moreover, the metric and the Riemann tensor are invariant under 
the global symmetry transformations of the group $G$, and their various components are strongly 
constrained. This simplifies the problem sufficiently much to enable us to solve it exactly. In addition 
to this fact, these spaces turn out to have constant curvature, since they are homogeneous. This 
suggests that the constraints emerging from the flatness and stability conditions will translate into 
particularly simple restrictions on the parameters of the theory. More precisely, the Riemann tensor 
in flat coordinates is a constant tensor that can be written in terms of a $G$-invariant combination 
of Kronecker $\delta$-functions that are invariant under the subgroup $H$:
\be
R_{I \bar J P \bar Q} = \mbox{combination of $\delta$-functions} \,.
\ee
The maps on Hermitian and symmetric tensors defined in Section 2 can be easily diagonalized 
in this case. Indeed, the eigentensors of these maps must correspond to irreducible representations 
of the group $H$, and can be obtained by decomposing the tensors $H^{I \bar J}$ and $S^{I J}$ under $G \to H$.

All the existing K\"ahler coset manifolds can be studied with the same technique. The form 
of the K\"ahler potential and the Riemann tensor with flat indices for all these spaces can 
be found in ref.~\cite{calabi}. Here we shall however restrict to those few spaces that are 
directly relevant for the simplest string models.

\subsection{\Large $\frac {SU(1,q+1)}{U(1)\times SU(q+1)}$}

The simplest class of K\"ahler coset manifold is the maximally symmetric space of dimension
$N=q+1$, with the structure:
\be
\label{cs}
{\cal M} = \frac {SU(1,q+1)}{U(1)\times SU(q+1)}\,.
\ee
This manifold is the K\"ahlerian analogue of the sphere in Riemannian geometry.
It can be parametrized by using a vector of complex fields $\phi_i$, where 
$i = 1,2,\dots,q+1$, and the K\"ahler potential is given by
\be
\label{k1}
K = -\, \frac 2{R_{\rm all}} \, {\rm ln}\, \Big(1 - \mbox{\large $\sum$}_i \Phi_i \Phi ^\dagger_i \Big)\,.
\ee
The Riemann tensor is in this case given by a tensor product of metrics, and in flat coordinates 
it has the simple form
\be
\label{decomp}
R_{I  \bar J  P \bar Q } = \frac {R_{\rm all}}2 \Big(\delta_{I \bar J}\, \delta_{P \bar Q}
+ \delta_{I \bar Q}\, \delta_{P \bar J}\Big)\,.
\ee

Let us now see what kind of information we can get by applying the general conditions (\ref{result1}) 
and (\ref{result3}) derived in Section 3. The map $H^{I \bar J} \rightarrow {R^{\bar J I}}_{\!\! P \bar Q} 
\,H^{P \bar Q}$ on Hermitian tensors does not have any vanishing eigenvalue, and can therefore be
inverted. One finds ${R^{-1\bar J I}}_{\!\! P \bar Q} = (2/R_{\rm all})(\delta^I_P \delta^{\bar J}_{\bar Q}
- (N \!-\! 1)^{-1} \delta^{I\bar J}\delta_{P \bar Q})$ and therefore the curvature constraint (\ref{result1}) 
implies that the overall curvature constant should satisfy $R_{\rm all} < 4/3 \, (q+1)/(q+2)$. 
On the other hand, the map $S^{I J} \rightarrow {R_{P \;\; Q}}^{\hspace{-10pt} I \;\;\, J} \,S^{P Q}$ on symmetric 
tensors has  eigenvalue $R_{\rm all}$ with degeneracy $(q+1)(q+2)/2$, so that the condition 
(\ref{result3}) implies the stronger bound $R_{\rm all} < 2/3$.

In this case, however, it is also possible to solve exactly the equations (\ref{condgen}), and 
compare it with the general conditions (\ref{result1}) and (\ref{result3}). To solve directly (\ref{condgen}),
we define the new positive and real variable 
\be
\label{comb1}
\Theta = \frac 1{\sqrt{3}}\, \sqrt{\mbox{\large $\sum$}_I \big|G^I\big|^2} \,.
\ee
The the two conditions (\ref{condgen}) can then be written as
\be
\label{ccc}
\Theta^2 = 1 \;,\;\; R_{\rm all} \,\Theta^4 < \frac 23 \,.
\ee
The situation is therefore identical to the one arising in a one-dimensional K\"ahler manifold 
with curvature $R_{\rm all}$. The constraint for the existence of non-supersymmetric flat and 
stable vacua is then simply 
\be
\label{condmax}
R_{\rm all}^{-1} > \frac 32 \,.
\ee
When this is satisfied, there is a unique solution corresponding to $\Theta = 1$. Note that we get the 
same result as the one obtained by using the, in principle, less restrictive condition (\ref{result3}). 
This illustrates the fact that it is possible to get useful information out of the conditions 
(\ref{result1}) and (\ref{result3}).

\subsection{\Large $\frac{SU(p,p+q)}{U(1)\times SU(p)\times SU(p+q)}$}

The next-to-simplest case of K\"ahler coset manifold is the Grassmanian 
space of dimension $N=p(p+q)$ given by:
\be
\label{gras}
{\cal M} = \frac {SU(p,p+q)}{U(1)\times SU(p) \times SU(p+q)}\,.
\ee
This is a natural and less symmetric generalization of the previous case, which is recovered for $p=1$. 
It can be parametrized by a matrix of complex fields $\phi_{i a}$, where $i=1,2,\dots,p$ and 
$a=1,2,\dots,p+q$. The K\"ahler potential is given by
\be
\label{k2}
K = -\, \frac 2{R_{\rm all}} \, {\rm ln}\,{\rm det} \Big(\delta_{i \barj}
- \mbox{\large $\sum$}_a \,\Phi_{i a} \Phi^\dagger_{j a} \Big)\,.
\ee
The Riemann tensor is in this case not given by a tensor product of metrics, unless $p=1$.
However, its components are nevertheless related in a simple way to those of the metric in certain
coordinate frames. In particular, in flat coordinates one finds the simple expression
\footnote{Note that the expression (\ref{r2}) can, also in this case, be rewritten in terms of the 
metric  in flat coordinates $\delta_{I A \, \bar J \bar B} = \delta_{I \bar J} \delta_{A \bar B}$,
but this decomposition takes a simple tensor product form as in (\ref{decomp}) only in 
the maximally symmetric case corresponding to $p=1$.} 
\cite{calabi,vhz}:
\be
R_{I A \, \bar J \bar B \, P C \, \bar Q \bar D} = \frac {R_{\rm all}}2
\Big(\delta_{I \bar J}\, \delta_{P \bar Q}\, \delta_{A \bar D}\, \delta_{C \bar B}
+ \delta_{I \bar Q}\, \delta_{P \bar J} \, \delta_{A \bar B} \, \delta_{C \bar D} \Big)\,.
\label{r2}
\ee

In this case, the map $H^{I A \, \bar J\bar B} \rightarrow {R^{\bar J \bar B \, I A}}_{P C \, \bar Q\bar D} 
\,H^{P C \, \bar Q \bar D}$ is singular and therefore not invertible, so that the condition (\ref{result1}) is 
trivially satisfied and does not give any constraint. The map 
$S^{I A\, J B} \rightarrow {R_{P C \hspace{12pt} Q D}}^{\hspace{-25pt} I  A \hspace{15pt} J B} \,S^{P C\,Q D}$ 
has, on the other hand, eigenvalues $R_{\rm all}$ with degeneracy $p(p+1)(p+q)(p+q+1)/4$
and $-R_{\rm all}$ with degeneracy $p(p-1)(p+q)(p+q-1)/4$. As this map has negative eigenvalues 
(unless $p=1$), the condition (\ref{result3}) is also satisfied and does not give any constraint 
either.

For these K\"ahler manifolds, nevertheless, one does find constraints by solving directly the equations 
(\ref{condgen}). To see this, notice that we can rewrite the conditions (\ref{condgen}) in a matrix form as 
${\rm tr} (G G^\dagger) = 3$ and $R_{\rm all}\, {\rm tr} (G G^\dagger G G^\dagger) < 6$. Observe now 
that the $p\times (p + q)$ matrix $G$ can be diagonalized by a bi-unitary transformation. More precisely, 
one can rewrite $G = U G^{\rm diag} V^\dagger$, where $U \in SU(p)$ and $V \in SU(p+q)$, and 
$G^{\rm diag}$ is a $p\times (p+q)$ diagonal matrix with $p$ complex eigenvalues. Note that, using 
this rewriting in the two conditions (\ref{condgen}) defining our problem, the matrices $U$ and $V$ always 
cancel out thanks to the cyclic property of the trace. Defining then the new $p$ positive and real variables
\be
\label{comb2}
\Theta_i = \frac 1{\sqrt{3}}\,\big|{\rm Eigenvalue}_i (G^{I A})\big| \,,
\ee
one can finally rewrite (\ref{condgen}) as:
\be
\label{4bis} 
\mbox{\large $\sum$}_i \, \Theta^2_i  = 1 \;,\;\;
\mbox{\large $\sum$}_i \, R_{\rm all} \, \Theta_i^4 < \frac 23 \,.
\ee
The problem takes now exactly the same form as the one for a factorizable scalar manifold given by the 
product of $p$ one-dimensional submanifolds all having the same curvature $R_i=R_{\rm all}$. 
The necessary condition for the existence of non-supersymmetric flat and stable vacua is then simply:
\be
R_{\rm all}^{-1} >\frac 3{2 p} \,.
\label{constraintRall}
\ee
If the curvature satisfies the restriction (\ref{constraintRall}), then there exist solutions, but 
only for a limited range of values for the variables $\Theta_i$. More precisely, one must have 
$\Theta_i \in [\Theta_{i}^-,\Theta_{i}^+]$ with:
\begin{eqnarray}
\begin{array}{lll}
\Theta_{i}^+ \!\!\!&=&\!\!\!
\left\{\!
\begin{array}{lll}
\displaystyle{
\sqrt{\frac 1p + \sqrt{\frac 23 \, \frac {p-1}{p}\Big(R_{\rm all}^{-1} - \frac 3{2p} \Big)}}} \,,
&\mbox{if}& R_{\rm all}^{-1} < \displaystyle{\frac 32} \,, \hspace{-30pt} \bigskip\ \\[-2mm]
1\,, &\mbox{if}&R_{\rm all}^{-1} >  \displaystyle{\frac 32} \,. \hspace{-30pt} \smallskip\ \\[-2mm]
\end{array}\right. \bigskip\ \\
\Theta_{i}^- \!\!\!&=&\!\!\!
\left\{\!
\begin{array}{lll}
\displaystyle{
\sqrt{\frac 1p - \sqrt{\frac 23 \, \frac {p-1}{p} \Big(R_{\rm all}^{-1} - \frac 3{2p} \Big)}}} \,,
&\mbox{if}& R_{\rm all}^{-1} < \displaystyle{\frac 3{2(p-1)}} \,, \hspace{-30pt} \bigskip\ \\[-2mm]
0\,, &\mbox{if}& R_{\rm all}^{-1} > \displaystyle{\frac 3{2(p-1)}} \,. \hspace{-30pt} \smallskip\ \\[-2mm]
\end{array}\right. \smallskip\ \\
\end{array}
\label{thetacritnbis}
\end{eqnarray}

\subsection{\Large $\frac{SO(2,q+2)}{SO(2)\times SO(q+2)}$}

Another simple and relevant kind of K\"ahler coset manifolds are the Grassmanian spaces 
of dimension $N = q+2$ of the form: 
\be
\label{grasreal}
{\cal M} = \frac {SO(2,q+2)}{SO(2) \times SO(q+2)}\,.
\ee
This coset manifold can be parametrized by a vector of complex fields $\phi_{i}$, with 
$i=1,2,\dots,q+2$, and a K\"ahler potential given by:
\be
\label{k3}
K = -\, \frac 2{R_{\rm all}} \, {\rm ln}\,\Big(1 - 2\, \mbox{\large $\sum$}_i  \Phi_i \Phi^\dagger_i
+\, \mbox{\large $\sum$}_{i,j} (\Phi_i \Phi^\dagger_j)^2 \Big)\,.
\ee
The Riemann tensor in flat coordinates is in this case found to have the slightly less trivial 
form \cite{calabi}:
\be
R_{I \bar J P \bar Q} = \frac {R_{\rm all}}2
\Big(\delta_{I \bar J}\, \delta_{P \bar Q} + \delta_{I \bar Q}\, \delta_{P \bar J} 
- \delta_{I P} \, \delta_{\bar J \bar Q} \Big)\,.
\label{r3}
\ee

In this case, the map $H^{I \bar J} \rightarrow {R^{\bar J I}}_{\!\! P \bar Q} \,H^{P \bar Q}$ is singular 
so that the condition (\ref{result1}) is satisfied and does not give any constraint. 
The map $S^{I J} \rightarrow {R_{P \;\; Q}}^{\hspace{-10pt} I \;\;\, J} \,S^{P Q}$ has, on the other hand, 
eigenvalues $R_{\rm all}$ with degeneracy $(q+1)(q+3)/2$
and $-\, q\,R_{\rm all}/2$ with degeneracy $1$. The condition (\ref{result3}) is therefore 
also satisfied and does not give any constraint either.

Nevertheless in this case one can also solve directly the conditions (\ref{condgen}) in an 
exact way. Using a vector notation, one can rewrite the conditions (\ref{condgen}) as $G \cdot G^* = 3$ 
and $2 (G \!\cdot\! G^*)^2 -  (G \!\cdot\! G) (G^* \!\cdot\! G^*) < 12/R_{\rm all}$. 
However now the two conditions depend only on two independent combinations 
of the variables. Indeed, introducing the two real positive variables
\be
\label{comb3}
\Theta_{1,2} = \frac 1{\sqrt{6}} \sqrt{\mbox{\large $\sum$}_I \big|G^I \big|^2 \pm
\sqrt{\Big(\mbox{\large $\sum$}_I \big|G^I \big|^2\Big)^2 - \Big|\mbox{\large $\sum$}_I \big(G^I \big)^2\Big|^2}} \,,
\ee
the two conditions can be rewritten as: 
\be
\label{4tris} 
\Theta^2_1 + \Theta^2_2 = 1 \;,\;\;
R_{\rm all} \, \Big(\Theta^4_1 + \Theta^4_2 \Big) < \frac 23 \,.
\ee  
The problem has now the same form as the one for a factorizable scalar manifold given by the 
product of two one-dimensional submanifolds with the same curvature $R_i=R_{\rm all}$. 
The necessary condition for the existence of non-supersymmetric flat and stable vacua is then:
\be
R_{\rm all}^{-1} >\frac 34 \,.
\label{constraintRallbis}
\ee
If the curvature satisfies the restriction (\ref{constraintRallbis}), then there exist solutions, but 
only for a limited range of values for the variables $\Theta_{1,2}$. More precisely, one must have 
$\Theta_{1,2} \in [\Theta_{1,2}^-,\Theta_{1,2}^+]$ with:
\begin{eqnarray}
\begin{array}{lll}
\Theta_{1,2}^+ \!\!\!&=&\!\!\!
\left\{\!
\begin{array}{lll}
\displaystyle{
\sqrt{\frac 12 + \sqrt{\frac {1}{3}\Big(R_{\rm all}^{-1} - \frac 34 \Big)}}} \,,
&\mbox{if}& R_{\rm all}^{-1} < \displaystyle{\frac 32}\,, \hspace{-30pt} \bigskip\ \\[-2mm]
1\,, &\mbox{if}& R_{\rm all}^{-1}> \displaystyle{\frac 32} \,. \hspace{-30pt} \smallskip\ \\[-2mm]
\end{array}\right. \bigskip\ \\
\Theta_{1,2}^- \!\!\!&=&\!\!\!
\left\{\!
\begin{array}{lll}
\displaystyle{
\sqrt{\frac 12 - \sqrt{\frac {1}{3} \Big(R_{\rm all}^{-1} - \frac 34 \Big)}}} \,,
&\mbox{if}& R_{\rm all}^{-1} < \displaystyle{\frac 32} \,, \hspace{-30pt} \bigskip\ \\[-2mm]
0\,, &\mbox{if}& R_{\rm all}^{-1} > \displaystyle{\frac 32} \,. \hspace{-30pt} \smallskip\ \\[-2mm]
\end{array}\right. \smallskip\ \\
\end{array}
\label{thetacritntris}
\end{eqnarray}

\section{Moduli spaces in string models}
\setcounter{equation}{0}

Many of the scalar manifolds arising in the moduli sector of string compactifications fall into the
classes of factorizable or symmetric spaces, actually homogeneous coset spaces, that we have
studied in the previous sections.
These sectors include the neutral fields controlling the size of the coupling and the geometry
of the compactification manifold, as well as possible Wilson lines for the hidden gauge group.
They represent natural candidates for the hidden sector in this type of models, and it is therefore
of evident interest to apply to these cases our results on the conditions under which flat and
locally stable non-supersymmetric vacua can exist.

In the simplest case of orbifold and orientifold compactifications, the untwisted sector moduli
space must be a subgroup of the moduli space emerging for maximally supersymmetric toroidal
reductions, which is uniquely fixed by the fact that there are $6$ extra internal dimensions
and by the rank $s$ of the hidden gauge group. More precisely, at leading order one finds:
\be
{\cal M} \subset \frac {SU(1,1)}{U(1)} \times \frac {SO(6,6+s)}{SO(6) \times SO(6+s)} \,.
\ee
The first factor is always present and is associated to the universal dilaton modulus $S$
controlling the coupling. The second factor is instead broken by the orbifold or orientifold
projection to a subgroup that has the form of a product of coset K\"ahler manifolds of the
types studied in Section 5. It is associated to the K\"ahler and complex structure moduli
$T$ and $U$ controlling the size and the shape of the compactification manifold,
and the Wilson lines $X$ of the hidden gauge group.

The simplest situation that can appear for a given modulus $\Phi_i$ is described by a K\"ahler
potential of the form \cite{logK}:
\be\label{kp1}
K = - \, n_i \, {\rm ln} \Big(\Phi_i + \Phi^\dagger_i \Big)\,.
\ee
It is straightforward to show that this potential can be written in the form (\ref{k1}), with $R_{\rm all} = 2/n_i$,
by means of a holomorphic change of variables and a K\"ahler transformation. The
corresponding K\"ahler manifold is therefore
\be
\label{s1}
{\cal M} = \frac {SU(1,1)}{U(1)}\,.
\ee
In this simplest case, the scalar manifold is both one-dimensional and symmetric, and
for the flatness and stability conditions this corresponds to having one field with curvature
$R_i = 2/n_i$. In the presence of several fields with K\"ahler potentials of the form (\ref{kp1}), 
the flatness and stability conditions imply $\sum_k R_k^{-1} > 3/2$, which requires that 
$\sum_{k}\, n_k > 3$, as found in ref.~\cite{us}.

A first relevant generalization involves the K\"ahler moduli controlling
the size of some cycle in the internal manifold, and is due to the possible presence of Wilson 
lines around that cycle. Each modulus $T_i$ can in principle mix with an arbitrary number $q_i$ 
of Wilson lines $X_{a_i}$, with $a_i = 1,2,\dots,q_i$. The simple one-dimensional space (\ref{s1}) 
is then enhanced to a $(q_i+1)$-dimensional space with a K\"ahler potential given by \cite{kounnas1}:
\be
\label{kk1}
K = -\, n_i \, {\rm ln}\,\Big(T_i + T_i^\dagger
- \mbox{\large $\sum$}_{a_i} X_{a_i} X^\dagger_{a_i} \Big)\,.
\ee
This K\"ahler potential (\ref{kk1}) can be written in the form (\ref{k1}), with $R_{\rm all} = 2/n_i$,
by means of a holomorphic field redefinition and a K\"ahler transformation. The corresponding
K\"ahler manifold is therefore isomorphic to a maximally symmetric coset space of the
type studied in subsection 5.1:
\be
\label{cs11}
{\cal M} = \frac {SU(1,q_i + 1)}{U(1)\times SU(q_i + 1)}\,.
\ee
Recalling the results of subsection 5.1, we conclude that, as far as the flatness and stability
conditions are concerned, this situation is equivalent to having a single field with curvature
$R_i = 2/n_i$.  In the presence of several groups of fields with K\"ahler potentials of the form (\ref{kk1}), 
the flatness and stability conditions imply then that $\sum_k R_k^{-1} > 3/2$, which requires again 
that $\sum_{k}\, n_k > 3$. Wilson lines do therefore not change qualitatively the
situation. Their presence enhances the minimal geometry in such a symmetric way that the
only effect of the corresponding auxiliary fields is to contribute together with the involved
modulus to the combination of auxiliary fields (\ref{comb1}) that is relevant to find the constraints.

Another interesting and relevant generalization can appear for K\"ahler moduli in particularly symmetric
models, like $\Z_2$ or $\Z_3$ orbifolds, and is due to the presence of additional non-standard moduli of
this type. More precisely, a set of $p_r$ K\"ahler moduli $T_{i_r}$ with equal parameter $n_r$, where
$i_r=1,2,\dots,p_r$, can mix with $p_r(p_r-1)$ extra K\"ahler moduli $T_{\alpha_r}$, where $\alpha_r=1,2,\dots,p_r(p_r-1)$.
There are then in total $p_r^2$ K\"ahler moduli, which can be organized in a matrix $T_{i_r j_r}$,
where $i_r,j_r=1,2,\dots,p_r$. The scalar manifold associated to the original $p_r$ moduli, which is
a product of $p_r$ copies of the minimal space (\ref{s1}), is then enhanced to a $p_r^2$-dimensional
space with a K\"ahler potential given by \cite{kounnas2}:
\be
\label{kk2}
K = -\, n_r \, {\rm ln}\,{\rm det} \Big(T_{i_r j_r} + T_{i_r j_r}^\dagger \Big)\,.
\ee
One can use a holomorphic field redefinition and a K\"ahler transformation to rewrite this
K\"ahler potential in the form (\ref{k2}), with $R_{\rm all} = 2/n_r$, and the corresponding
scalar manifold is therefore a particular complex Grassmanian manifold of the type studied in 
subsection 5.2:
\be
\label{cs22}
{\cal M} =  \frac {SU(p_r, p_r)}{U(1)\times SU(p_r) \times SU(p_r)}\,.
\ee
From the analysis developed in subsection 5.2 we can then conclude that the flatness and stability
conditions depend only on $p_r$ independent combinations of fields (\ref{comb2}) with identical 
curvatures $R_r = 2/n_r$. In the presence of several groups of fields with K\"ahler potentials of the form 
(\ref{kk2}), the flatness and stability conditions imply that $\sum_{r}\, p_r\,R_r^{-1} > 3/2$, which reduces
to the condition $\sum_r p_r\, n_r > 3$. The extra off-diagonal K\"ahler moduli are therefore 
qualitatively irrelevant for the restrictions imposed on the curvature, and they just combine with the 
diagonal K\"ahler moduli into the combinations of fields (\ref{comb2}) relevant to find the 
constraints.

The two deformations that we have considered so far, related to the presence of extra Wilson lines
and off-diagonal K\"ahler moduli, can also occur simultaneously. In this more general situation, a
set of $p_r$ K\"ahler moduli $T_{i_r}$ with equal parameter $n_r$, where $i_r=1,2,\dots,p_r$,
can mix with $p_r(p_r-1)$ extra K\"ahler moduli $T_{\alpha_r}$, where $\alpha_r=1,2,\dots,p_r(p_r-1)$,
as well as $p_r q_r$ Wilson lines $X_{i_r a_r}$, where $a_r = 1,2,\dots,q_r$. There are then
$p_r^2$ K\"ahler moduli, which can be organized in a matrix $T_{i_r j_r}$, and in addition
$p_r q_r$ Wilson lines $X_{i_r a_r}$. The scalar manifold associated to the original $p_r$ moduli,
which is a product of $p_r$ copies of the minimal space (\ref{s1}), is then enhanced to a
$p_r(p_r+q_r)$-dimensional space with a K\"ahler potential given by \cite{kounnas2}:
\be
\label{kk3}
K = - \, n_r \, {\rm ln}\,{\rm det} \Big(T_{i_r j_r} + T_{i_r j_r}^\dagger
- \mbox{\large $\sum$}_{a_r} X_{i_r a_r} X^\dagger_{j_r a_k}\Big)\,.
\ee
This can be shown to be equivalent to a K\"ahler potential of the form (\ref{k2}) with $R_{\rm all} = 2/n_r$,
and the corresponding K\"ahler manifold is now the general case of the complex Grassmanian
manifolds studied in subsection 5.2:
\be
\label{cs33}
{\cal M} =  \frac {SU(p_r, p_r + q_r)}{U(1)\times SU(p_r) \times SU(p_r + q_r)}\,.
\ee
As in the previous case we can use the information given in subsection 5.2 to conclude that the flatness
and stability conditions depend only on $p_r$ independent combinations of fields (\ref{comb2}) with 
identical curvatures $R_r = 2/n_r$. In the presence of several groups of fields with K\"ahler
potentials of the form (\ref{kk3}), the flatness and stability conditions imply that 
$\sum_r p_r R_r^{-1} > 3/2$, which requires as in the previous cases that $\sum_r p_r\, n_r > 3$. 
This means that neither the extra off-diagonal K\"ahler moduli nor the extra matter fields $X_{i_r a_r} $ 
are relevant for the restrictions imposed on the curvature, and they just combine with the diagonal K\"ahler 
moduli into the combinations of fields (\ref{comb2}) relevant to find the constraints.

There is yet another type of interesting enhancement that can appear for K\"ahler and complex structure
moduli in certain specific models, like $\Z_2 \times \Z_2$ orbifolds, and which is due to Wilson lines
mixing with these two kinds of fields simultaneously. More precisely, a pair of two K\"ahler and complex
structure moduli $T_r$ and $U_r$ associated to the same submanifold and having the same parameter
$n_r$ can mix with a number $q_r$ of Wilson lines $X_{a_r}$, with $a_r=1,2,\dots,q_r$, associated to
that submanifold. The scalar manifold associated to the two original moduli, which is a product of two
copies of the minimal space (\ref{s1}), is then enhanced to a $(q_r+2)$-dimensional space with a
K\"ahler potential given by \cite{JP2}:
\be
\label{kk5}
K = -\, n_r \, {\rm ln} \, \Big((T_r + T^\dagger_r) (U_r + U^\dagger_r)
- \mbox{\large $\sum$}_{a_r} (Z_{a_r} \!+ Z^\dagger_{a_r})^2\Big)\,.
\ee
In this case, the scalar manifold can be shown to be
\be
\label{cs55}
{\cal M} = \frac {SO(2,2+q_r)}{SO(2) \times SO(2+q_r)}\,.
\ee
This is a Grassmanian coset of the type described in subsection 5.3. We can then conclude that the
flatness and stability conditions depend only on the two independent combinations of fields
(\ref{comb3}) with identical curvatures $R_r = 2/n_r$. In the presence several such pairs of fields 
with K\"ahler potentials of the form (\ref{kk5}), the flatness and stability conditions imply
$\sum_r 2 \, R_r^{-1} > 3/2$, which requires that $\sum_r 2\,n_r > 3$. The presence of the
Wilson lines is once again qualitatively irrelevant, and just changes the relevant combinations
(\ref{comb3}) of fields.

Summarizing we have learned that all the possible enhancements we have considered
of the minimal factorized moduli space given by a product of factors (\ref{s1}) do not 
qualitatively change the form of the flatness and stability conditions. Due to the very high 
degree of symmetry of these enhancements, the only net effect of the extra fields involved 
is to change the combinations of fields that are relevant for the conditions. In particular,
they do not change neither the number of relevant combinations nor the values of the 
associated curvatures. The curvature constraints for the existence of flat and stable 
non-supersymmetric vacua then depend only on the number $N$ of diagonal moduli 
$\Phi_i$ and their associated numerical parameter $n_i$. The problem is thus identical 
to the one arising in a $N$-dimensional factorizable space with constant curvatures given 
by $R_i = 2/n_i$. In all the situations analyzed here, the existence of non-supersymmetric 
flat and stable vacua is then permitted only if the parameters $n_i$ fulfill the condition
\be
\mbox{\large $\sum$}_k \,n_k > 3 \,.
\ee
The only peculiarity of situations where the moduli space is enhanced is that some 
of the $n_i$'s have the same values.

In the simplest situations arising in string compactifications, there are seven moduli fields.
These are just the $S$ field, the three K\"ahler moduli $T_i$ and the three complex structure 
moduli $U_i$ which in the simplest situations have diagonal potential of the form (\ref{kp1}), 
or more in general we can have mixtures of these fields with the extra moduli fields enhancing 
the K\"ahler manifold spanned by them. Each of these fields has $n_i = 1$, and this means 
that none of them can dominate on its own supersymmetry breaking, as was already pointed 
out in ref.~\cite{us}. In fact, as can be easily seen from (\ref{thetacritn}), the variables $\Theta_i$ 
have an upper and a lower bound given by:
\be
\Theta_i^\pm = \sqrt{\frac {n_i \pm \sqrt{n_i \big(\sum_k n_k - n_i \big)\big(\sum_k n_k - 3\big)}}{\sum_k n_k}} \,.
\ee
This constrains as well the values that the $F$ auxiliary fields associated to each of the independent 
combinations of fields can take, which depend both on the diagonal and the off-diagonal moduli that 
might be present.

\section{Radion in Randall--Sundrum models}
\setcounter{equation}{0}

The general results we have derived in the previous sections have also interesting implications 
for phenomenological models with a single extra dimension, like for instance supersymmetric 
Randall--Sundrum models with generic warping $k$ \cite{RS}. In such models, the classical 
effective K\"ahler potential has a form that is constrained by locality and general covariance. 
Denoting by $M_5$ the $5$-dimensional Planck scale, by $T$ the radion chiral multiplet controlling 
the size of the extra dimension, and by $X_a$, $a=1,2,\dots,q$, and $\tilde X_{\tilde a}$, 
$a=1,2,\dots,\tilde q$, the matter fields at the two branes, one finds that \cite{RSeffective1,RSeffective2}:
\be
\label{KRS}
K = - \, 3 \, {\rm ln} \bigg[\frac {M_5^3}k \Big(1 - e^{- k (T + T^\dagger)}\Big)
- \frac 13 \, \mbox{\large $\sum$}_a X_a^\dagger X_a - \frac 13 \, \mbox{\large $\sum$}_{\tilde a}
\tilde X_{\tilde a}^\dagger \tilde X_{\tilde a} \, e^{- k (T + T^\dagger)} \bigg] \,.
\ee
It is straightforward to show, by means of a simple rescaling of the fields, that this
describes a $(q + \tilde q + 1)$-dimensional maximally symmetric coset space
of the form:
\be
\label{Msym}
{\cal M} = \frac {SU(1,q + \tilde q + 1)}{U(1) \times SU(q+\tilde q + 1)} \,.
\ee
From (\ref{KRS}) we can read that the curvature is given by $R_{\rm all} = 2/3$, and therefore,
using the result given in (\ref{condmax}), it marginally violates the curvature bound allowing for the
existence of flat and stable non-supersymmetric vacua. This means in particular that corrections to the
K\"ahler potential (\ref{KRS}) will be crucial, since even a slight change on the curvature can allow it to 
fulfill the necessary condition (\ref{condmax}).

This fact has strong implications on models of radius stabilization within this setup. For instance
in the model proposed in ref.~\cite{LSuplift,RSeffective2} the radion superpotential induced
by some gaugino condensation in the bulk is used to stabilize the radion at an AdS point,
which is then uplifted to a Minkowski vacuum thanks to a brane sector. Using our previous result, 
we conclude that this model cannot work at leading order if the brane sector sector has a K\"ahler 
potential as in (\ref{KRS}). To improve the situation one needs to have some non-linearity in the matter
sector (this would reduce the high degree of symmetry of (\ref{Msym})), like for instance a non-trivial
field-dependent wave-function factor. This was already suggested in ref.~\cite{b2b}, where it
was assumed that such a wave-function would stabilize the brane scalar fields at vanishing values. The
results derived here show that this is actually mandatory in order for this model to work.

In general, quantum effects induce non-trivial corrections to the tree-level K\"ahler potential
(\ref{KRS}). These corrections can be either divergent local effects that can be reabsorbed by a
renormalization of the parameters in (\ref{KRS}), or finite non-local effects that induce, on the contrary,
a correction with a different dependence on the fields (see for instance refs.~\cite{b2b,Adam}).
These Casimir-like corrections modify the structure of (\ref{KRS}), and can therefore be potentially 
useful to lower the effective curvature below the critical value $R_{\rm all}= 2/3$ obtained at the 
classical level.

\section{Conclusions}
\setcounter{equation}{0}

In this paper, we have analyzed in more generality the implications of the flatness and stability 
constraints derived in ref.~\cite{us} for supergravity theories where only chiral multiplets are relevant 
for supersymmetry breaking. We have explored in detail special cases where the K\"ahler manifolds 
spanned by the scalar fields are such that it is possible to work out in full generality the implications
of these constraints. We have studied in particular the coset K\"ahler manifolds that are relevant for the
moduli sector of string models. Since these are homogeneous spaces with constant
curvature, the implications of the constraints are in this case particularly simple and directly
related to the parameters of the theory. We have found that the conditions for the existence of flat and 
stable non-supersymmetric vacua impose in these cases strong constraints on the K\"ahler geometry 
and also on the values that the auxiliary fields can take (as was also the case for the examples 
considered in \cite{us}). We were able to show that the basic symmetry enhancements due to the addition 
of extra off-diagonal and/or untwisted matter fields that extend the minimal space of products of 
$SU(1,1)/U(1)$ factors to more complicated coset manifolds are qualitatively irrelevant as far as the 
constraints on the K\"ahler geometry are concerned. Actually, the additional fields were found to 
change only the combinations of auxiliary fields that are relevant for the constraints, leaving their 
number and the associated constraints unchanged. We have also explored the case of completely 
arbitrary scalar manifolds, for which the original variational problem defined by the constraints is 
not exactly solvable. Nevertheless, we were able to derive explicit results also in this more
general case by further simplifying the conditions defining the problem in a way that made the 
new variational problem exactly solvable. In this way, we were able to obtain weaker but completely
general necessary conditions.

There are many avenues of future work following these lines. One of the most interesting is to perform 
the the same kind of analysis as the one presented here when also vector multiplets participate to 
supersymmetry breaking. The presence of vector multiplets with significant $D$ auxiliary fields, 
in addition to chiral multiplets with non-vanishing $F$ auxiliary fields, can alleviate the restrictions 
found in ref.~\cite{us} and in this paper (where $D$-terms where neglected with respect to 
$F$-terms) \footnote{A toy example where this is the case can be found in ref.~\cite{vz}}.
More precisely, some of the vector multiplets can gauge some isometries of the chiral multiplet
sector, and the corresponding $D$-terms are then related to the $F$-terms through the Killing
potentials specifying the gauging. The progress made in the present paper concerning symmetric
spaces should be relevant to study this interesting but more complicated situation more efficiently.

\section*{Acknowledgments}
\setcounter{equation}{0}
We thank M.~Blau, J.-P.~Derendinger,  E.~Dudas, S.~Ferrara, M.~Petropoulos, R.~Rattazzi and 
F.~Zwirner for valuable suggestions and discussions. This work has been partly supported by the 
Swiss National Science Foundation and by the European Commission under contracts 
MRTN-CT-2004-005104. We also thank the Theory Division of CERN for hospitality.

\end{document}